\documentclass[a4paper,twocolumn,11pt,accepted=2026-03-18]{quantumarticle}
\pdfoutput=1
\usepackage[utf8]{inputenc}
\usepackage[english]{babel}
\usepackage[T1]{fontenc}
\usepackage{amsmath}

\usepackage{tikz}
\usepackage{lipsum}

\usepackage{makeidx}
\usepackage[utf8]{inputenc}
\usepackage{graphicx}
\usepackage{mathrsfs}
\usepackage{graphicx}
\usepackage{braket}
\usepackage[subrefformat=parens,labelformat=parens,caption=false]{subfig}
\usepackage{amsfonts}
\usepackage{amssymb}
\usepackage{amsthm}
\usepackage{amsmath}
\usepackage{mleftright}
\usepackage{dcolumn}
\usepackage{color}
\usepackage[dvipsnames]{xcolor}
\usepackage{verbatim}
\usepackage{cancel}
\usepackage{multirow}
\usepackage[breaklinks,
            colorlinks,
            urlcolor=blue,
            linkcolor=blue,
            anchorcolor=blue,
            citecolor=blue]{hyperref}
           
\usepackage{soul}

\newcommand\Tr{\text{Tr}}

\newcommand{\dmtwo}[1]{{\color{black} #1}}

\usepackage[numbers,sort&compress]{natbib}

\newcommand{\parens}[1]{\left(#1\right)}
\newcommand{\sparens}[1]{\left[ #1 \right]}

\newcommand{\an}[1][]{\hat{a}^{#1}}
\newcommand{\ad}[1][]{\hat{a}^{\dag #1}}
\newcommand{\bigO}[1]{\mathcal{O}\mleft(#1\mright)}

\newcommand{\SM}{Appendix}

\begin{document}

\title{Power-law distributions in nonequilibrium open quantum systems}

\author{Wai-Keong Mok}
\email{darielmok@caltech.edu}
\affiliation{Institute for Quantum Information and Matter, California Institute of Technology, Pasadena, CA 91125, USA}
\begin{abstract}
    Power-law probability distributions are widely used to model extreme statistical events in complex systems, with applications to a vast array of natural phenomena ranging from earthquakes to stock market crashes to pandemics. We show that analogous heavy tails arise naturally in open quantum systems with nonlinear dissipation. Introducing a prototypical family of quantum dynamical models, we analytically prove the emergence of power-law tails in the steady state energy distribution, originating from an amplification of quantum noise whose microscopic fluctuations grow with energy. 
    Moreover, our analysis suggests a general mechanism for heavy-tail statistics in the nonequilibrium steady states of open quantum systems: Nonlinear dissipation generically induces multiplicative quantum noise, enforced by the constraints of quantum mechanics, which is responsible for the heavy-tail behavior. This is supported by numerical simulations of a general class of nonlinear dynamics known as quantum Li\'{e}nard systems. Remarkably, even when the corresponding classical system is stable, we find power-law tails in both steady-state populations and coherences, which occur for typical parameters without fine-tuning. This phenomenon can potentially be harnessed to develop extreme photon sources for novel applications in light-matter interaction and sensing.
\end{abstract}
\maketitle

\section{Introduction}
The natural occurrence of power-law probability distributions underpins many fascinating statistical phenomena in complex systems across vast domains such as physics, biology, geology, economics and social science~\cite{newman2005power,marquet2005scaling,corral2019power,gabaix2009power}. Power-law distributions are often associated with the presence of extreme or `black swan' events, which are statistical events that occur with low frequency but have enormous impact~\cite{carla2012review,markovic2014power}. This is sometimes stated in various contexts as the \textit{Pareto principle}: a majority of the observed effects arise from a minority of the causes. Extreme events abound in nature, with well-known examples including earthquakes~\cite{sornette1996rank}, rogue ocean waves~\cite{dudley2019rogue}, solar flares~\cite{cliver2022extreme}, stock market crashes~\cite{gabaix2003theory} and pandemics~\cite{machado2020rare}. 

Often, extreme events are observed as emergent phenomena in large-scale systems, and statistical models are constructed to explain the empirical data \textit{a posteriori}. Here, we show that such a phenomenon can emerge dynamically in arguably one of the most fundamental stochastic systems in nature --- a single dissipative quantum system. While extreme events and heavy-tail statistics (such as power laws) are well-studied in classical systems, their occurrences in quantum-mechanical systems have been largely unexplored. To this end, we construct the \textit{$M$-boson model}, a prototypical toy model of a quantum dynamical system with nonlinear dissipation, and prove analytically that the energy distribution in the steady state exhibits a power-law tail. If we regard the system physically as a quantum particle trapped in a potential well, this phenomenon manifests as a power-law decay in the position and momentum distribution of the quantum particle~\cite{lillo2000anomalous,batle2002quantum}.

The physical origin of the power-law distribution in our model can be attributed to quantum noise~\cite{gardiner2000quantum}. The phenomenon stems from the intimate connection between microscopic fluctuations and dissipation in quantum mechanics, and traces back to Heisenberg's uncertainty principle. From the perspective of open quantum systems, the microscopic interactions between the system and the external bath, which are responsible for the dissipation, unavoidably inject noise into the system~\cite{gardiner2000quantum,breuer2007theory,carmichael1999statistical}. Furthermore, the form of this quantum noise is highly dependent on the type of dissipation. For nonlinear dissipation, this noise is typically multiplicative, which means the noise is coupled to the quantum state of the system~\cite{chia2020phase,chia2023quantum}. This coupling induces a self-amplification of the noise, in which the scale of the fluctuations grows with the excitation of the system. The amplified fluctuations at large excitation are only weakly suppressed by the dissipation, and the net result is a power-law distribution.

Our analysis suggests a general mechanism for power laws to emerge in the nonequilibrium steady states of open quantum systems describing quantum particles subject to dissipative forces, arising from the aforementioned multiplicative quantum noise. We provide numerical evidence for this hypothesis by studying the open quantum systems obtained via quantizing a general classical nonlinear dynamical system~\cite{chia2025quantization}. Of particular interest is the Li\'{e}nard system, a general class of dynamical systems encompassing many paradigmatic examples such as the van der Pol, Rayleigh, and Duffing oscillators~\cite{strogatz2024nonlinear}. Analogous to their classical counterparts, quantum Li\'{e}nard systems have been established to exhibit limit cycles~\cite{chia2020relaxation,arosh2021quantum,shen2023quantum,chia2025quantization}, and display interesting nonlinear phenomena such as synchronization, relaxation oscillations, and excitable dynamics. We numerically demonstrate a new phenomenon unique to the quantum system: for typical parameters in the classically stable regime (without fine-tuning), the steady state density matrix of quantum Li\'enard systems exhibits power-law tails in both the populations (i.e., diagonal elements) and the coherences (i.e., off-diagonal elements) in the Fock basis, highlighting the essential role of quantum noise.

Interestingly, both the $M$-boson and quantum Li\'enard models can exhibit a regime where the power-law tail vanishes so slowly such that the quantum state, while normalizable, contains an infinite number of excitation quanta on average. Moreover, the distribution for the excitation number is highly skewed towards low excitation. This gives rise to extreme events in which the system, when measured in the excitation number basis, produces highly excited states containing orders of magnitude more excitation quanta than the typically measured values. This regime of the power law was previously observed experimentally in a different nonlinear quantum optical setup, which works by amplifying the vacuum fluctuations~\cite{manceau2019indefinite}. In that experiment, extreme emission events containing $\sim 10^6$ photons in a single pulse were observed.

In the context of quantum optics, where the excitations are photons, our analytical results show that the system can exhibit an extreme case of photon superbunching in which the emitted photons are significantly more correlated in time as compared to thermal light sources~\cite{mcneil1975possibility}. The intensity correlation function (at zero delay) in this case can be made arbitrarily high, due to the strong fluctuations in the electromagnetic field. This phenomenon of power-law behavior opens up potential avenues for exploring extreme regimes of light-matter interactions~\cite{heimerl2024multiphoton,stammer2023quantum,gorlach2023high}, and can be a useful resource for sensing applications such as ghost imaging~\cite{gatti2004ghost,chan2009high}.

\section{$M$-boson model}
\begin{figure}[h]
\centering
\subfloat{%
  \includegraphics[width=0.99\linewidth]{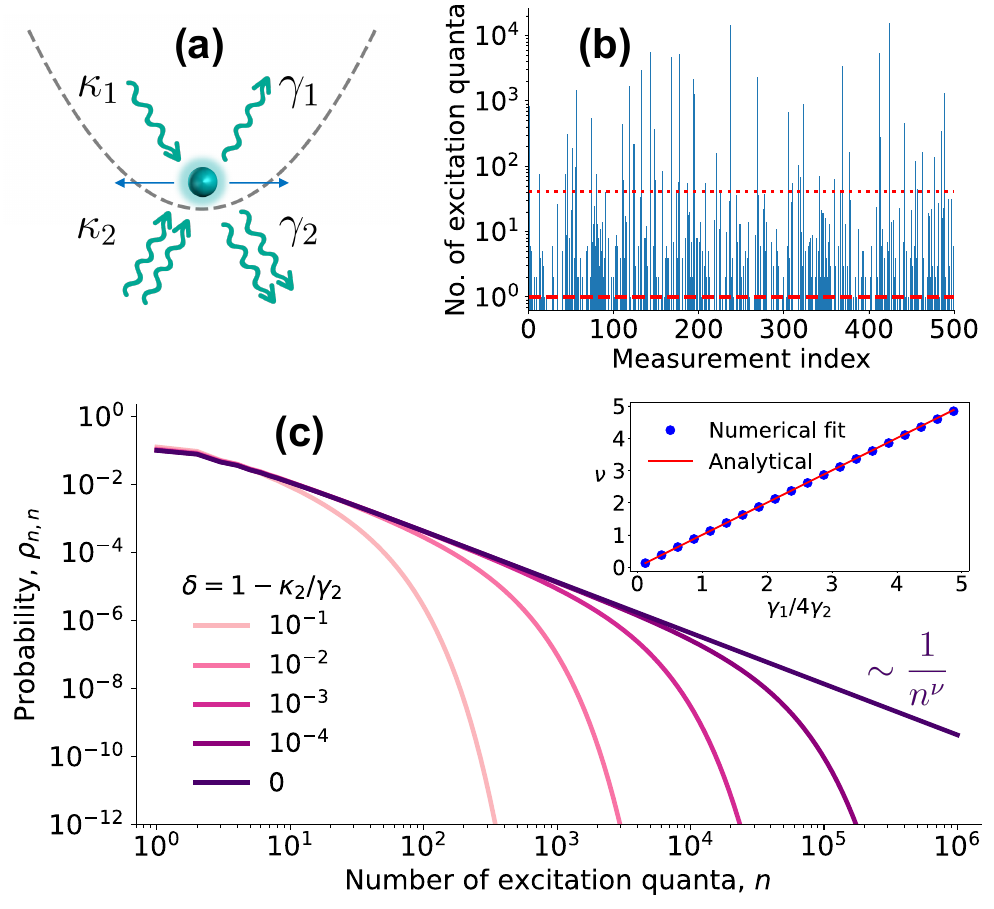}%
}
\caption{\textbf{Power-law distribution in the $M$-boson model}. (a) A quantum particle trapped in a one-dimensional harmonic potential interacts with $M$ independent baths ($M = 2$ depicted). The particle exchanges $m$ excitation quanta with the $m$-th bath ($m = 1,\ldots,M$), at a dissipation rate of $\gamma_m$ and pumping rate of $\kappa_m$. (b) Simulated measurement for the number of excitation quanta $n$ on the steady state of Eq.~\eqref{eq:Mphoton_mastereq} with $M = 2$ and $500$ measurement samples. The dashed line indicates the median count $n = 1$, and the dotted line indicates the $90^{\text{th}}$ percentile count $n \approx 41.1$. Extreme events reaching $n \approx 16000$ are observed. Parameters are $\{\gamma_1,\kappa_1,\gamma_2,\kappa_2\} = \{6,0,1,1\}$. (c) The probability distribution $\rho_{n,n}$ for $n$ exhibits a power-law tail $\sim n^{-\nu}$ at the critical point $\kappa_2 = \gamma_2$, plotted for $\gamma_1/\gamma_2 = 6$ and $\kappa_1/\gamma_2 = 0$. The inset shows the power-law exponent $\nu$ against $\gamma_1/4\gamma_2$. The data points are obtained by numerically fitting with a power law, which agrees excellently with the analytical result ~\eqref{eq:powerlawtail}, $\nu = \gamma_1/4\gamma_2$. Away from the critical point, the probability distribution follows a power law up to an exponential cutoff at $n \sim M/\delta$.}
 \label{fig:Mboson}
\end{figure}
We begin by analyzing, in detail, a minimal toy model which exhibits the power-law tail. Consider the system to be a 1D quantum harmonic oscillator, with a characteristic energy scale $\hbar \omega_0$. The system interacts with $M$ independent baths by exchanging $m$ excitation quanta with the $m$-th bath ($m = 1,\ldots,M$), at dissipation and pumping rates denoted by $\gamma_m$ and $\kappa_m$ respectively [see Fig.~\ref{fig:Mboson}(a)]. This can be achieved by preparing the $m$-th bath at the resonant frequency $m \omega_0$. For a bath at thermal equilibrium, the ratio $0 \leq \kappa_m/\gamma_m \leq 1$ is determined by the temperature of the bath, where the lower and upper limits correspond to zero and infinite temperatures respectively. Under the Born-Markov approximation, the quantum dynamics of the system can be modelled by the Lindblad master equation~\cite{gardiner2000quantum,breuer2007theory,carmichael1999statistical}
\begin{equation}
    \dot{\rho} = -i[\omega_0 \ad \an, \rho] + \sum_{m=1}^{M} \parens{\gamma_m \mathcal{D}[\an[m]]\rho + \kappa_m \mathcal{D}[\ad[m]]\rho}
\label{eq:Mphoton_mastereq}
\end{equation}
for the density matrix $\rho$ of the system, where $\mathcal{D}[\hat{c}]\rho \equiv \hat{c} \rho \hat{c}^\dag - (\hat{c}^\dag \hat{c} \rho + \rho \hat{c}^\dag \hat{c})/2$ is the dissipator. The annihilation and creation operators $\an$ and $\ad$ for the system excitation satisfy the bosonic commutation relation $[\an,\ad] = 1$. The model can physically describe, for instance, photonic (in optical or microwave cavities) or phononic excitations (in a trapped ion or nanomechanical resonator). In Eq.~\eqref{eq:Mphoton_mastereq}, the bath degrees of freedom are traced out, and the effects of the microscopic system-bath interactions are captured by the dissipator terms. A microscopic derivation of the $M = 2$ case can be found in Refs.~\cite{simaan1978off,gilles1993two,chia2020phase}; the derivation for $M > 2$ follows analogously. We shall refer to Eq.~\eqref{eq:Mphoton_mastereq} as the \textit{$M$-boson model}. The case of $M = 1$ reduces to the well-known example of a damped harmonic oscillator~\cite{gardiner2000quantum,breuer2007theory,carmichael1999statistical}, while the $M = 2$ case, known as the quantum Stuart-Landau oscillator~\footnote{This is also commonly referred to as the quantum van der Pol oscillator in the literature, e.g., Refs.~\cite{walter2014quantum,walter2015quantum}. However, since the corresponding classical model is the Stuart-Landau model (which describes limit-cycle oscillators near the Hopf bifurcation), this model should be more appropriately called the quantum Stuart-Landau model.}, was previously studied in the context of quantum limit cycles and synchronization~\cite{walter2014quantum,walter2015quantum,lorch2016genuine,mok2020synchronization,chia2023quantum}. We remark that the two-boson ($\gamma_2$) dissipation term was implemented experimentally on superconducting circuits using dissipation engineering, to stabilize Schr\"{o}dinger cat states for bosonic quantum error correction~\cite{leghtas2015confining,lescanne2020exponential}, and also recently in trapped ions~\cite{li2025experimental,liu2025observation} to demonstrate quantum synchronization. A similar setup can be used to implement the $\kappa_2$ pumping term.

Since Eq.~\eqref{eq:Mphoton_mastereq} is rotationally symmetric (i.e., invariant under the $\text{U}(1)$ transformation $\an \to e^{i\phi} \an$), the steady state density matrix is diagonal in the excitation number basis $\{\ket{n}\}$ and independent of $\omega_0$. In this basis, Eq.~\eqref{eq:Mphoton_mastereq} simplifies to
\begin{equation}
\begin{split}
    \dot{\rho}_{n,n} &= \sum_{m=1}^{M} \gamma_m \sparens{\rho_{n+m,n+m}\prod_{j=1}^{m}(n+j) - \rho_{n,n} \prod_{j=0}^{m-1} (n-j)} \\&+ \sum_{m=1}^{M} \kappa_m \sparens{\rho_{n-m,n-m} \prod_{j=0}^{m-1}(n-j) - \rho_{n,n}\prod_{j=1}^{m}(n+j)},
\label{eq:diagonal_EOM}
\end{split}
\end{equation}
with the coherences (off-diagonal elements of $\rho$) decoupled from the probability distribution $\rho_{n,n}$. Solving for the steady state $\dot{\rho}_{n,n} = 0$ yields the truncated power-law tail
\begin{equation}
\rho_{n,n} \propto \frac{e^{-n \delta / M}}{n^{\nu}}, \quad 
    \nu = \frac{M-1}{M^2} \frac{\gamma_{M-1} - \kappa_{M-1}}{\gamma_M}
\label{eq:powerlawtail}
\end{equation}
valid for large $n$, small $\delta = 1 - \kappa_M/\gamma_M \geq 0$, and any $M \geq 2$ (see \SM~\ref{app:Mboson_steadystate} for details). Since detailed balance is generally not satisfied, this describes a nonequilibrium steady state. The power-law tail is exact at the critical point $\delta = 0$, which can be physically interpreted as fixing the $M$-th bath at infinite temperature. The power-law behavior is almost entirely controlled by the leading order processes involving $M$ and $M-1$ bosons, which is intuitive since these processes dominate the dynamics at large $n$. By tuning the rates $\gamma_{M-1}$ and $\kappa_{M-1}$ (for example, by adjusting the temperature of the $(M-1)$-th bath), the power-law exponent $\nu$ can vary between $0$ and $\infty$. The regime $\nu \leq 2$ is particularly interesting, where the mean excitation number is infinite at criticality. As shown in Fig.~\ref{fig:Mboson}(b), this can give rise to extreme events where the measured excitation numbers are orders of magnitude higher than typical values, following the power-law behavior displayed in Fig.~\ref{fig:Mboson}(c). More generally, the $k$-th moment of the number distribution $\rho_{n,n}$ is infinite or undefined if $\nu \leq k+1$. Away from the critical point, Eq.~\eqref{eq:powerlawtail} predicts an exponential cutoff at $n \sim M/\delta$. For $\delta \ll 1$, the power-law behavior holds for a large range of $n$. This suggests that the dramatic effects of the power-law tail, such as extreme events, can be observed close to the critical point. Such a scenario is also more realistic than the infinite-temperature limit $\delta = 0$, where physical limitations in the setup or measurement apparatus introduce a natural high-energy cutoff in $n$.

\begin{figure*}
\centering
\subfloat{%
  \includegraphics[width=0.99\linewidth]{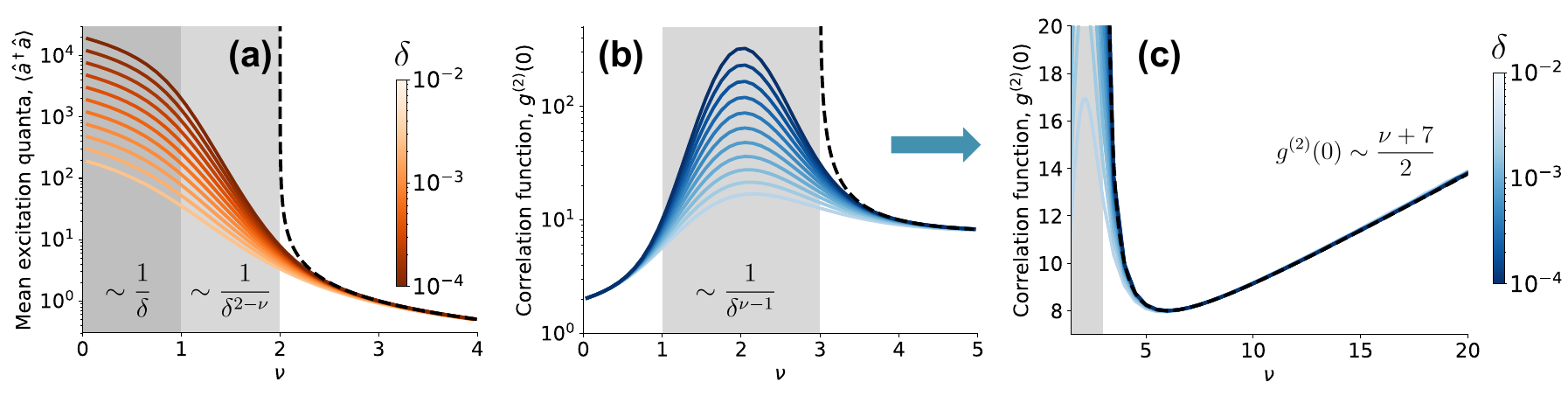}%
}
\caption{\textbf{Physical signatures of the power-law distribution}. Numerical results for the $M$-boson model with $M = 2$ and $\kappa_1 = 0$. (a) Mean number of excitation quanta $\braket{\ad \an}$, (b, c) Normalized second-order correlation function $g^{(2)}(0)$. $\delta = 1 - \kappa_2/\gamma_2$ is the deviation from the critical point. The power-law exponent is $\nu = \gamma_1/4\gamma_2$ [Eq.~\eqref{eq:powerlawtail}]. In all the plots, the solid curves are obtained numerically for $\delta \in [10^{-4},10^{-2}]$. The Hilbert space truncation is set at $N_T = 5 \times 10^5$ to ensure numerical convergence. The black dashed lines indicate the corresponding analytical results~\eqref{eq:meanphoton} and~\eqref{eq:g20}. Grey shaded regions denote the regime in which the physical quantity is divergent near the critical point, with the divergence derived from the scaling theory indicated. At large $\nu$, $g^{(2)}(0)$ scales linearly with $\nu$.}
 \label{fig:physicalsignatures}
\end{figure*}

The power-law behavior at criticality might appear surprising, since the $M$-boson exchanges in and out of the system occur at precisely the same rate $\gamma_M = \kappa_M$, and classical intuition suggests that there is no net flow of excitations. The emergence of the power law can be understood by examining the quantum fluctuations. Taking the white-noise limit of the bath, the master equation~\eqref{eq:Mphoton_mastereq} yields, in the Heisenberg picture, the quantum It\^{o} equation~\cite{gardiner2000quantum,hudson1984quantum,chia2020phase,chia2023quantum}
\begin{equation}
\begin{split}
    d\an &= dt \sum_{m=1}^{M} \frac{m}{2} \parens{-\gamma_m \ad[m-1] \an[m] + \kappa_m\an[m] \ad[m-1]} \\
    &+ \sum_{m=1}^{M} m \ad[m-1] d\hat{W}_m.
\end{split}
\label{eq:itoSDE_general}
\end{equation}
The final term showcases the quantum noise explicitly, via the independent operator-valued Wiener increments $d\hat{W}_m$ which satisfy $d\hat{W}_m d\hat{W}_{m^\prime}^\dag = \delta_{mm^\prime} \gamma_m dt$ and $d\hat{W}_m^\dag d\hat{W}_{m^\prime} = \delta_{mm^\prime} \kappa_m dt$. The noise operators are necessary to preserve the bosonic commutation relation $[\hat{a}(t),\hat{a}^\dag(t)] = 1$ (in the Heisenberg picture). This is in sharp contrast with classical stochastic systems, where the noise terms can be introduced quite freely by hand (at least from a mathematical standpoint). Crucially, the quantum noise is additive only for $m = 1$, and multiplicative for all $m > 1$. In the latter case, the noise operator $d\hat{W}_m$ is coupled to the bosonic field $\ad[m-1]$. This causes the fluctuations to be self-amplified and grow rapidly, which must be counterbalanced by the damping forces for the steady state to exist. At large excitation numbers, the dominant damping term in Eq.~\eqref{eq:itoSDE_general} is $\propto (\gamma_M - \kappa_M) \ad[M-1] \an[M]$, which vanishes at the critical point $\kappa_M/\gamma_M = 1$. The remaining damping forces can only weakly suppress the fluctuations at large $n$, which results in the power-law tail.

Our analysis reveals that the heavy-tail statistics of the $M$-boson model can be attributed to the multiplicative quantum noise in Eq.~\eqref{eq:itoSDE_general}, which arises naturally from the nonlinear dissipation terms $\mathcal{D}[\hat{a}^m]\rho$ for $m > 1$. More generally, for an arbitrary Lindblad operator $\hat{c}$, the noise term in Eq.~\eqref{eq:itoSDE_general} is of the form $\sim [\hat{c}^\dag,\an]d\hat{W} + [\an,\hat{c}]d\hat{W}^\dag$~\cite{hudson1984quantum}, which is multiplicative if $\hat{c}$ is nonlinear in $\an$ or $\ad$. This suggests a mechanism for heavy-tail statistics to emerge in the steady state in more general nonequilibrium open quantum systems:
\begin{enumerate}
    \item Quantum dissipation is invariably associated with quantum noise (this is true of any open quantum system).
    \item Crucially, the form of the quantum noise is not arbitrary. It is heavily constrained by the type of dissipation, in order to preserve commutation relations, e.g., $[\hat{a}(t),\hat{a}^\dag(t)] = 1$, where $\hat{a}(t)$ is the bosonic annihilation operator in the Heisenberg picture.
    \item For nonlinear dissipation, the quantum noise is typically multiplicative. The competition between multiplicative noise (which tends to cause amplification) and nonlinear dissipation (which tends to suppress large excitations) gives rise to the heavy-tail (e.g., power-law) statistics in the steady state.
\end{enumerate}
To support this argument, we will numerically study a general class of dynamical systems known as quantum Li\'{e}nard systems~\cite{chia2025quantization} in Sec.~\ref{sec:lienard}, and show that power-law tails appear not only in the steady state populations $\rho_{n,n}$, but also in the steady-state coherences $\rho_{n,m}$ ($m \neq n$). Before doing so, we first discuss some measurable signatures of the power law.

\section{Physical signatures of the power-law tail} The power-law distribution can be directly observed in experiments by measuring the system in the excitation number basis, such as by photodetection, if the bosonic operator $\an$ represents an electromagnetic mode. This was achieved experimentally in Ref.~\cite{manceau2019indefinite}, using a different physical system based on nonlinear optics. There, the origin of the power law is also attributed to amplified quantum noise. In this section, we explore several other physical properties in which the power-law behavior manifests. For simplicity, we will illustrate this using the $M$-boson model due to its analytical tractability. However, note that much of the discussion here holds more generally for any quantum state exhibiting the power-law tail $\rho_{n,n} \sim n^{-\nu}$. We restrict to $M = 2$, which already captures most of the essential physics, while analogous results for $M > 2$ are discussed in the Appendix.

At the critical point $\delta = 0$, the mean occupation number in the steady state is exactly
\begin{equation}
    \braket{\ad \an} = \frac{1+\kappa_1/(4\gamma_2)}{\nu - 2},
\label{eq:meanphoton}
\end{equation}
where $\nu = (\gamma_1 - \kappa_1)/4\gamma_2$ is the power-law exponent, see \SM~\ref{app:moments_number_distr}. Equation~\eqref{eq:meanphoton} is valid only for $\nu > 2$, and is plotted in Fig.~\ref{fig:physicalsignatures}(a). The predicted divergence at $\nu = 2$ is consistent with the power-law tail~\eqref{eq:powerlawtail}. For $\nu < 2$, the mean excitation number $\sim \exp[(2-\nu)4\gamma_2 t]$ blows up exponentially in time. The second moment can also be computed in the steady state as
\begin{equation}
    \braket{\ad[2] \an[2]} = \braket{(\ad \an)^2} - \braket{\ad \an} = \frac{1 + (8+\kappa_1/\gamma_2)\braket{\ad \an}}{2(\nu - 3)}
\label{eq:number_moment2}
\end{equation}
for $\nu > 3$, and blows up exponentially $\sim \exp[(3-\nu)8\gamma_2 t]$ for $\nu < 3$. Again, the divergence at $\nu = 3$ is consistent with the power-law tail. All higher moments can be computed recursively, as derived in \SM~\ref{app:moments_number_distr}.

A useful quantity to characterize the bosonic field is the normalized second-order correlation function $g^{(2)}(0) \equiv \braket{\ad[2] \an[2]}/\braket{\ad \an}^2$, plotted in Figs.~\ref{fig:physicalsignatures}(b,c). In the context of an electromagnetic field, $g^{(2)}(0)$ indicates the likelihood of two photons arriving simultaneously at a detector, with $g^{(2)}(0) = 1$ for coherent light described by Poissonian statistics~\cite{mandel1995coherence}. Incoherent light typically exhibits bunching, i.e., $g^{(2)}(0) > 1$, such as thermal light sources which have $g^{(2)}(0) = 2$. The second-order correlation function can be evaluated from Eqs.~\eqref{eq:meanphoton} and~\eqref{eq:number_moment2} for $\nu > 3$ (setting $\kappa_1 = 0$ for simplicity),
\begin{equation}
    g^{(2)}(0) = \frac{(\nu - 2)(\nu + 6)}{2(\nu - 3)}.
\label{eq:g20}
\end{equation}
This has a minimum value of $g^{(2)}(0) = 8$ at $\nu = 6$, which reflects the fact that the steady state is not at thermal equilibrium. $g^{(2)}(0)$ diverges at $\nu = 3$ due to the strong fluctuations in the bosonic field. For $\nu \gg 1$, $g^{(2)}(0) \sim (\nu+7)/2$ grows linearly with $\nu$ [see Fig.~\ref{fig:physicalsignatures}(c)]. This regime could be accessible in practical implementations where $\{\gamma_2,\kappa_2\} \ll \gamma_1$. Thus, the linear behavior of $g^{(2)}(0)$ can potentially serve as an experimental signature of the power-law tail. We remark that the $k$-boson correlation function analogous to Eq.~\eqref{eq:g20} scales as $g^{(k)}(0) \equiv \braket{\ad[k] \an[k]}/\braket{\ad \an}^k \sim \nu^{\left \lfloor{k/2}\right \rfloor}$ for large $\nu$, see \SM~\ref{app:correlation_func}.

The behavior of $g^{(2)}(0)$ for $\nu \leq 3$ is less straightforward to analyze. Obviously, the correlation function is infinite in the regime $2 < \nu \leq 3$, since $\braket{\ad \an}$ is finite, while $\braket{\ad[2] \an[2]}$ diverges. However, for $\nu \leq 2$, both $\braket{\ad[2] \an[2]}$ and $\braket{\ad \an}$ diverge at the critical point. We can gain insights about the physics for $\nu < 2$ using a simple scaling theory (see \SM~\ref{app:scaling} for details): Suppose $\rho_{n,n} \propto n^{-\nu}$ has a sharp cutoff at $n_c = 1/\delta$. A straightforward calculation predicts the scaling behavior $\braket{(\ad \an)^k} \sim 1/\delta^k$ for $\nu \in (0,1)$ and $\sim 1/\delta^{k-\nu+1}$ for $\nu \in (1,k+1)$, with $k \in \mathbb{Z}^+$. Consequently, $g^{(2)}(0)$ diverges as $1/\delta^{\nu-1}$ for $\nu \in (1,2)$. Replacing the sharp cutoff with the exponential cutoff in Eq.~\eqref{eq:powerlawtail} only adds a finite correction and does not affect the scaling of the divergences. The predicted divergences from the scaling theory agree with numerical simulations near the critical point, as shown in Fig.~\ref{fig:physicalsignatures}(b). Note that the observed behavior in Fig.~\ref{fig:physicalsignatures}(b) cannot be extrapolated to the critical point $\delta = 0$, since $g^{(2)}(0) \sim \exp(8\gamma_2 t)$ blows up exponentially in time for all $\nu < 2$. Mathematically, this corresponds to a singular perturbation. In the regime $\nu \ll \delta$, the steady state is close to thermal equilibrium and thus $g^{(2)}(0) \approx 2$, consistent with the numerical observation.

The power-law behavior can also manifest in the quantum phase space. Introducing the dimensionless quadrature operators $\hat{q} = (\an + \ad)/\sqrt{2}$ and $\hat{p} = -i(\an - \ad)/\sqrt{2}$, with the canonical commutation relation $[\hat{q},\hat{p}] = i$, we show analytically in \SM~\ref{app:wigner} that the Wigner quasiprobability distribution in the steady state has the asymptotic behavior
\begin{equation}
    W(q,p) \sim \frac{1}{(q^2 + p^2)^{\nu}}
\label{eq:wigner}
\end{equation}
with the same power-law exponent $\nu$ as the number distribution. This can be observed, for example, by performing tomography on the bosonic mode. If the system represents a quantum particle confined in a one-dimensional harmonic potential, the position and momentum distributions of the particle exhibit the tail behavior $\sim |q|^{1-2\nu}$ and $\sim |p|^{1-2\nu}$ respectively, from Eq.~\eqref{eq:wigner}.

To summarize, the power-law distribution due to quantum fluctuations gives rise to extreme events which can be orders of magnitude above the typical measured values. As shown above, such effects can be observed through various physical quantities that have a scaling behavior dependent on the power-law exponent $\nu$. We emphasize that the behavior of the various physical quantities shown in Fig.~\ref{fig:physicalsignatures} is not intrinsic to the $M$-boson model. A similar behavior is expected for any quantum system exhibiting a power-law tail $\rho_{n,n} \sim n^{-\nu}$, although the value of $\nu$ may be model-dependent. 

\section{Power-law behavior in general nonlinear dynamics}
\label{sec:lienard}
The $M$-boson model serves as a toy model to gain insights about the physical origin of the power-law tail. However, one might argue that the phenomenon is not truly quantum mechanical, since Eq.~\eqref{eq:diagonal_EOM} is a classical master equation, and multiplicative noise is well-known to produce power-law distributions in classical systems~\cite{biro2005power,sornette1998multiplicative}. Thus, the more interesting question is: Are there dissipative quantum systems that exhibit power-law tails in the steady state not only in the populations, but also in the coherences (i.e., off-diagonal elements of $\rho$)? If so, this would indicate that the intrinsically quantum nature of the noise plays an important role.

To this end, let us consider a general classical dynamical system of the form
\begin{equation}
    \frac{d^2 q}{dt^2} = F(q,\dot{q}),
\end{equation} where $F$ is an arbitrary function describing the force acting on a classical particle. We then construct its quantization, i.e., a Lindblad master equation that yields the generalized Ehrenfest theorem
\begin{equation}
    \frac{d\braket{\hat{q}}}{dt} = \braket{\hat{p}}, \quad \frac{d\braket{\hat{p}}}{dt} = \braket{F(\hat{q},\hat{p})}, 
\end{equation}
with some suitably defined operator ordering. Here, $\hat{q}$ and $\hat{p}$ are the quadrature operators previously defined. For the evolution to be physically valid, we impose the constraint that the master equation must generate dynamical maps that are completely positive and trace preserving (CPTP)~\cite{lindblad1976on,breuer2007theory}. This general quantization problem was recently solved in Ref.~\cite{chia2025quantization}, for arbitrary bivariate polynomials $F(q,p)$. A rich subclass of dynamical systems is the cubic Li\'enard system~\cite{strogatz2024nonlinear}, with the quantum equation of motion given by 
\begin{equation}
\begin{split}
    \frac{d\braket{\hat{q}}}{dt} &= \braket{\hat{p}}, \\ \frac{d\braket{\hat{p}}}{dt} &= k_0\braket{\hat{p}}- k_2 \braket{:{\hat{q}^2 \hat{p}}:} - k_1 \braket{\hat{q}} - k_3 \braket{:\hat{q}^3:}.
\end{split}
\label{eq:quantum_lienard}
\end{equation}
The notation $\braket{: \hat{O} :}$ indicates normal ordering, after expressing the operator $\hat{O}$ in terms of $\an$ and $\ad$. The choice of operator ordering here is simply for convenience, and does not affect the discussion~\cite{chia2025quantization}. The cubic Li\'enard system encompasses paradigmatic nonlinear systems such as the van der Pol, Rayleigh, and Duffing oscillators~\cite{strogatz2024nonlinear}. Physically, the parameters $k_2$ and $k_3$ represent the nonlinear damping and nonlinear frequency response (i.e., anharmonicity) of the system, respectively. In the absence of the nonlinearity, i.e., $k_2 = k_3$ = 0, the cubic Li\'{e}nard system reduces to the damped harmonic oscillator. 
\begin{figure}
\centering
\subfloat{%
  \includegraphics[width=0.99\linewidth]{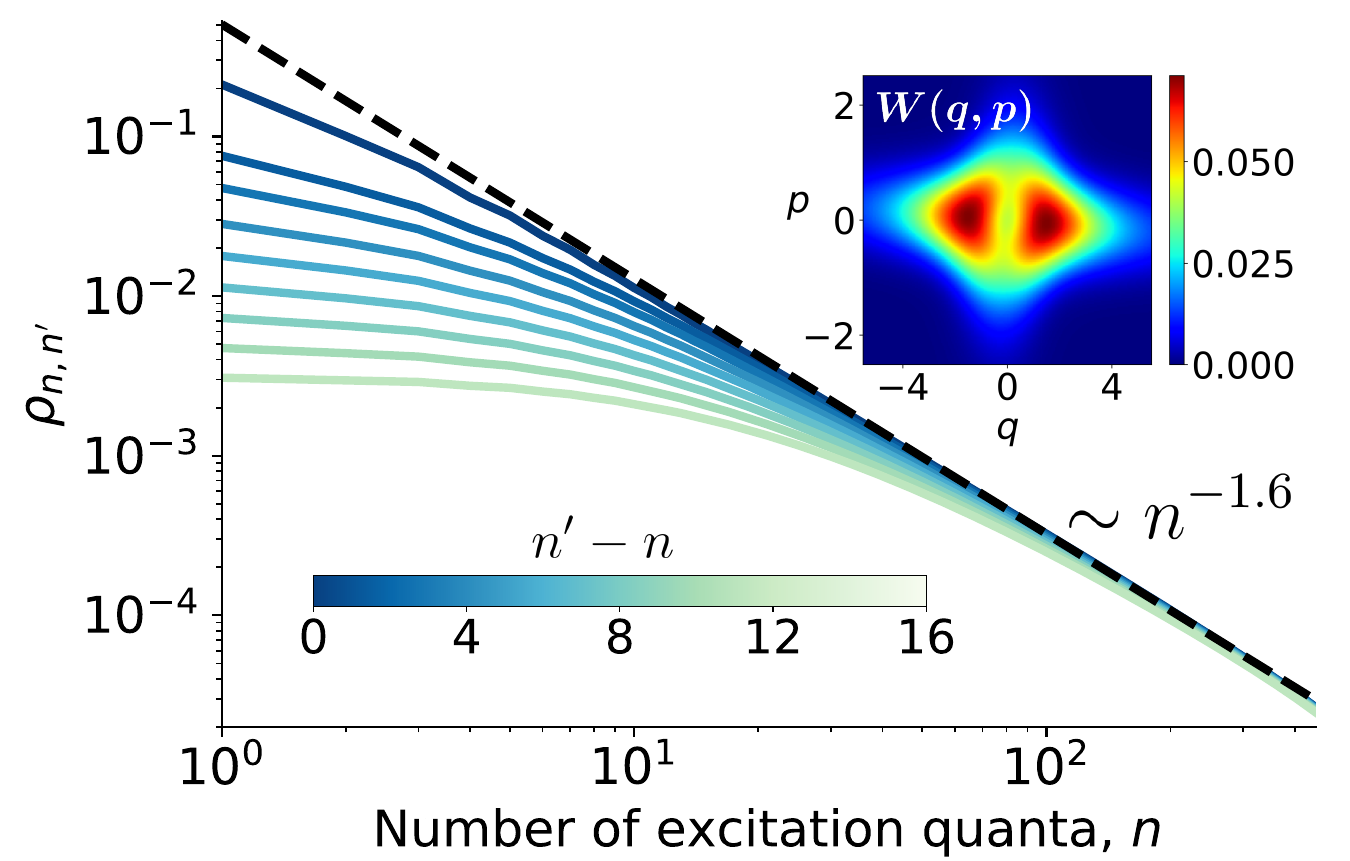}%
}
\caption{\textbf{Power-law tails in the quantum Li\'{e}nard system}. Magnitudes of steady-state density matrix element $|\rho_{n,n^\prime}|$ against the number of excitation quanta $n$ for the quantum Li\'{e}nard system~\eqref{eq:quantum_lienard} in the limit-cycle regime. Curves plotted are for even $n^\prime - n \in [0,16]$. Matrix elements for odd $n^\prime - n$ are identically zero. The black dashed line shows the power-law scaling $|\rho_{n,n^\prime}| \propto n^{-1.6}$ obtained from fitting the number distribution $\rho_{n,n}$. (Inset) Wigner quasiprobability distribution $W(q,p)$ of the steady state. $W(q,p)$ is mostly supported near the phase space origin, with a heavy tail that decays slowly. Parameters: $(k_0,k_1,k_2,k_3) = (0.169,0.12,1,0.035)$, with a Hilbert space truncation of $N_T = 10^3$ in the simulation for numerical convergence. Similar behavior can be observed for other parameters.}
 \label{fig:lienard}
\end{figure}
Using the quantization method developed in Ref.~\cite{chia2025quantization}, we obtain the Lindblad master equation corresponding to Eq.~\eqref{eq:quantum_lienard},
\begin{equation}
\begin{aligned}
    \dot{\rho} = &-i[\hat{H},\rho] + k_0 \mathcal{D}[\ad]\rho + \frac{k_2}{2} \mathcal{D}[\ad \an - \ad[2]/2]\rho \\ &+ \frac{3k_2}{8} \mathcal{D}[\an[2]]\rho,
\end{aligned}
\end{equation}
with the Hamiltonian
\begin{equation}
    \hat{H} = \hat{H}_0 + \hat{H}_1,
\end{equation}
\begin{equation}
    \hat{H}_0 = \frac{k_1+1}{2} \ad \an + \frac{3 k_3}{8} \ad[2] \an[2]
\end{equation}
and
\begin{equation}
\begin{aligned}
    \hat{H}_1 &= \frac{i}{4}\sparens{k_0 - i(k_1 - 1)} \an[2] - \frac{i}{4} \parens{\frac{k_2}{2} + i k_3} \ad \an[3] \\&- \frac{i}{16}(k_2 + i k_3) \an[4] + \text{h.c.},
    \end{aligned}
\end{equation}
where $\text{h.c.}$ denotes the Hermitian conjugate of all the preceding terms.

Although the quantum models are constructed phenomenologically, they can potentially be realized using quantum dissipation engineering (also known as reservoir engineering)~\cite{poyatos1996quantum,kienzler2015quantum,harrington2022engineered}, which has been widely successful for implementing nonlinear dissipation~\cite{leghtas2015confining,lescanne2020exponential,lorch2017quantum,aiello2022quantum,li2025experimental,liu2025observation,venkatraman2024nonlinear,lorch2016genuine}. The key idea is to engineer a nonlinear Hamiltonian interaction between the system and a highly dissipative ancillary system, which mediates an effective nonlinear dissipation in the system~\cite{lescannethesis}. A wide range of nonlinear dissipation terms can be implemented this way in circuit QED, using asymmetrically threaded SQUIDs (ATS) or similar architectures~\cite{lescanne2020exponential,berdou2023one,vanselow2026dissipating,hua2025engineering}. Alternatively, one can use continuous measurement and feedback to realize the nonlinear dissipation~\cite{wiseman2009quantum}, which was proposed to implement a variant of Eq.~\eqref{eq:quantum_lienard_general}~\cite{chia2020relaxation}. 

A key qualitative difference with the $M$-boson model~\eqref{eq:Mphoton_mastereq} is that the dynamical system~\eqref{eq:quantum_lienard} does not possess continuous rotational symmetry. Therefore, the steady state (if it exists) will contain coherences in the excitation number basis. In the regime of weak nonlinear dissipation $k_2 \ll k_1$, the master equation can be described by a quantum Stuart-Landau model with a generalized squeezing drive~\cite{sonar2018squeezing, shen2023quantum}, which has an exponentially vanishing number distribution (see Appendix~\ref{app:quantum_lienard}). Thus, strong nonlinear dissipation is needed to observe the power-law tail. 

In the classical case, for parameters $k_i > 0$, $i = 0,1,2$ and $k_3 \geq 0$, Li\'{e}nard's theorem guarantees a unique stable limit cycle surrounding the origin~\cite{strogatz2024nonlinear}. Correspondingly, quantum limit cycles also exist in the quantum system~\cite{chia2020relaxation,arosh2021quantum,shen2023quantum}. We provide numerical evidence for the emergence of power-law tails in the quantum steady state. Unlike the $M$-boson model~\eqref{eq:Mphoton_mastereq}, the power law in the number distribution $\rho_{n,n}$ can be observed without fine-tuning of the parameters. We demonstrate this by simulating the master equation with parameters $k_0, k_1$ and $k_3$ randomly sampled from $[0,1]$ (in units of $k_2$). An explicit example is shown in Fig.~\ref{fig:lienard}, with additional numerical results presented in Appendix~\ref{app:lienard_additional_numerics}. Interestingly, the power law emerges not only in the probabilities $\rho_{n,n}$, like for the $M$-boson model, but also in the off-diagonal elements (coherences) of the density matrix. In the example shown, we observe the scaling $|\rho_{n,n^\prime}| \sim n^{-1.6}$ for $n^\prime - n$ up to $16$. This is in the regime of infinite-mean excitation number, akin to $\nu < 2$ in the $M$-boson model. This example suggests that the quantum effects of the fluctuations are relevant. \dmtwo{A qualitatively similar behavior is observed for $k_0 = k_1 = k_3 = 0$ and $k_2 = 1$, see Appendix~\ref{app:lienard_additional_numerics}. This supports the intuition that the nonlinear dissipation term governed by $k_2$ is the key mechanism responsible for the emergence of power-law behavior.} While beyond the scope of this work, a systematic analysis of the quantum fluctuations in such models would be valuable. This can potentially provide an analytical explanation for the observed power laws in our numerical simulations, and elucidate how the power-law exponent depends on the parameters. 

\section{Discussion}
In this work, we propose the emergence of power-law distributions and extreme events as a novel feature of nonlinear quantum dissipation. This can be understood from the fact that the quantum noise associated with nonlinear dissipation is typically multiplicative in nature. That is, the fluctuations are coupled to the quantum state and become amplified at higher excitation. The competition between dissipation and noise amplification results in the power-law behavior. We introduce the prototypical $M$-boson model which is analytically proven to produce power-law energy distributions at the critical point. The ubiquity of this phenomenon is further supported via numerical simulations on a general class of nonlinear dynamical systems which are quantum analogs of Li\'{e}nard systems. We show that power laws can also emerge in the (off-diagonal) coherences of the density matrix, which exhibit the same scaling as the (diagonal) probability distribution.

Conceptually, the emergence of power-law tails also calls into question how classical dynamics arise from quantum dissipative systems. In standard examples such as the damped harmonic oscillator and the quantum Stuart-Landau model~\cite{walter2014quantum,mok2020synchronization}, the quantum fluctuations are relatively insignificant at large amplitudes (which can be formalized using the system-size expansion~\cite{carmichael1999statistical}), giving a well-defined classical limit. On the contrary, the presence of power-law tails originating from multiplicative quantum noise implies that the quantum fluctuations are not negligible in such cases, even at large amplitudes. Reconciling nonlinear dynamical systems with open quantum systems thus remains an interesting open problem. It would also be interesting as a future theoretical study to connect the results of this work with known mechanisms of the power law~\cite{newman2005power}, such as scale invariance, universality and self-organized criticality.

\textit{Acknowledgements.---} I would like to thank Tobias Haug, Jessica Jiang, Jingu Pang, John Preskill, Gil Refael and Yu Tong for insightful discussions. I am also particularly grateful to Andy Chia, Leong-Chuan Kwek and Changsuk Noh for collaborations on related projects. The Institute for Quantum Information and Matter is an NSF Physics Frontiers Center. 

\bibliographystyle{unsrtnat}
\bibliography{bib}

\newpage
\appendix
\onecolumngrid

\section{Steady state number distribution for the $M$-boson model}
\label{app:Mboson_steadystate}
Consider the $M$-boson model, with the Lindblad master equation
\begin{equation}
    \dot{\rho} = -i[\omega_0 \ad \an, \rho] + \sum_{m=1}^{M} \parens{\gamma_m \mathcal{D}[\an[m]]\rho + \kappa_m \mathcal{D}[\ad[m]]\rho}.
\label{eq:Mboson_ME_supp}
\end{equation}
This describes a quantum harmonic oscillator subject to incoherent processes such as damping (governed by the $\gamma_m$ terms) and pumping (governed by the $\kappa_m$ terms). On top of the harmonic oscillator Hamiltonian $H_0 = \hbar \omega_0 \ad \an$, one can also consider a generalized model by adding any Hamiltonian which is diagonal in the number (Fock) basis $\{ \ket{n} \}$. This includes, for example, a Kerr nonlinearity $H_{\text{Kerr}} = K \ad[2] \an[2]$ which controls the anharmonicity of the oscillator. Such terms do not contribute to the dynamics of the populations (diagonal elements of $\rho$ in the number basis), 
\begin{equation}
\begin{split}
    \dot{\rho}_{n,n} &= \sum_{m=1}^{M} \gamma_m \sparens{\rho_{n+m,n+m}\prod_{j=1}^{m}(n+j) - \rho_{n,n} \prod_{j=0}^{m-1} (n-j)} \\&+ \sum_{m=1}^{M} \kappa_m \sparens{\rho_{n-m,n-m} \prod_{j=0}^{m-1}(n-j) - \rho_{n,n}\prod_{j=1}^{m}(n+j)},
\end{split}
\end{equation}
as given in the main text. For $M = 2$, the exact solution to the steady state populations can be found in Ref.~\cite{arosh2021quantum}. However, the solution is a complicated expression in terms of hypergeometric functions, and is cumbersome to work with. Here, we are interested in the asymptotic behavior of $\rho_{n,n}$ for large $n$. To this end, we propose the truncated power-law ansatz
\begin{equation}
\label{eq:exponential_powerlaw_ansatz}
    \rho_{n,n} \propto \frac{e^{-n/\xi}}{n^{\nu}},
\end{equation}
for large $n$, where $\xi$ and $\nu$ are to be determined. This gives the relation
\begin{equation}
    \rho_{n+m,n+m} = \rho_{n,n} e^{-m/\xi} \parens{1+\frac{m}{n}}^{-\nu} = \rho_{n,n} e^{-m/\xi} \parens{1 - \frac{\nu m }{n} + \frac{\nu(\nu - 1) m^2}{2 n^2} + \ldots}
\label{eq:pops_relation}
\end{equation}
between the populations. Substituting Eq.~\eqref{eq:pops_relation} into the steady state condition $\dot{\rho}_{n,n} = 0$, we obtain
\begin{equation}
\begin{split}
    0 &= \sum_{m=1}^{M} \gamma_m n^m \sparens{ e^{-m/\xi} \parens{1+\frac{m(m+1-2\nu)}{2n} + \ldots} - \parens{1-\frac{m(m-1)}{2n} + \ldots}} \\
    &+ \sum_{m=1}^{M} \kappa_m n^m \sparens{ e^{m/\xi} \parens{1 - \frac{m(m-1- 2\nu)}{2n} + \ldots} - \parens{1+\frac{m(m+1)}{2n} + \ldots}}.
\label{eq:exponents_relation}
\end{split}
\end{equation}
To leading order $\bigO{n^M}$, we have
\begin{equation}
    \gamma_M \parens{e^{-M/\xi} - 1} + \kappa_M \parens{e^{M/\xi} - 1} = 0
\end{equation}
which can be readily solved to yield
\begin{equation}
    \xi = \frac{M}{\ln(\gamma_M/\kappa_M)}.
\end{equation}
Let us write $\delta = 1 - \kappa_M/\gamma_M$, where $\delta \geq 0$ is required for the steady state to exist. Near the critical point $\delta = 0$, we can write
\begin{equation}
    \xi = \frac{M}{-\ln(1-\delta)} \approx \frac{M}{\delta}.
\label{eq:xi_approx}
\end{equation}
The parameter $\xi$ controls the onset of the exponential cutoff, which occurs roughly at $n \sim \xi \approx M/\delta$. As $\delta \to 0^+$, the exponential cutoff becomes irrelevant, and the number distribution $\rho_{n,n} \sim n^{-\nu}$ exhibits a power-law tail. To obtain the power-law exponent $\nu$, we consider Eq.~\eqref{eq:exponents_relation} to the subleading order $\bigO{n^{M-1}}$, which gives
\begin{equation}
\begin{split}
    &\frac{\gamma_M}{2} \parens{e^{-M/\xi} M(M+1-2\nu) + M(M-1)} + \gamma_{M-1} \parens{e^{-(M-1)/\xi}-1} \\
    -& \frac{\kappa_M}{2} \parens{e^{M/\xi} M(M-1-2\nu) + M(M+1)} + \kappa_{M-1} \parens{e^{(M-1)/\xi} - 1} = 0.
\end{split}
\end{equation}
For small $\delta$, we can solve for $\nu$ approximately to get the power-law exponent
\begin{equation}
\label{eq:powerlawexponent}
    \nu \approx \frac{M-1}{M^2} \frac{\gamma_{M-1} - \kappa_{M-1}}{\gamma_M - \kappa_M} \delta = \frac{M-1}{M^2} \frac{\gamma_{M-1} - \kappa_{M-1}}{\gamma_M}
\end{equation}
as stated in the main text. To summarize, near the critical point, the number distribution has a truncated power-law of the form
\begin{equation}
\boxed{
\rho_{n,n} \propto \frac{e^{-n \delta/M}}{n^{\nu}},
}
\label{eq:truncatedpowerlaw}
\end{equation}
where $\nu$ is given in Eq.~\eqref{eq:powerlawexponent}. At $\delta = 0$, the exponential cutoff vanishes, and the distribution exhibits a power-law tail for arbitrarily large $n$. 

It is straightforward to show that the steady state does not generically obey detailed balance, and is therefore nonequilibrium. Detailed balance imposes the constraint
\begin{equation}
    \rho_{n,n} \kappa_m \prod_{j=1}^{m}(n+j) = \rho_{n+m,n+m}\gamma_m \prod_{j=0}^{m-1}(n-j),
\end{equation}
for all $m = 1,\ldots,M$. Substituting the asymptotic behavior~\eqref{eq:exponential_powerlaw_ansatz} for $\rho_{n,n}$, and matching both sides to leading order in $n$, yields the constraint
\begin{equation}
    \xi = \frac{m}{\ln(\gamma_m/\kappa_m)}.
\end{equation}
This is only satisfied for $m = M$, and not generally for other values of $m$.

\section{Analytical results for $M = 2$ at the critical point}
\label{app:critical_point_M2}
Here, we present additional analytical results for the $M$-boson model, in the simplest non-trivial case of $M = 2$.

\subsection{Moments of the number distribution}
\label{app:moments_number_distr}
From the master equation for the $M$-boson model, the equation of motion for a general observable $\hat{O}$ is given by
\begin{equation}
    \frac{d\braket{\hat{O}}}{dt} = \Tr\parens{\hat{O}\dot{\rho}} = i\omega_0 \braket{[\ad \an, \hat{O}]} + \sum_{m=1}^{M} \frac{\gamma_m}{2} \parens{ \braket{\ad[m] [\hat{O},\an[m]]} + \braket{[\ad[m],\hat{O}]\an[m]} } + \frac{\kappa_m}{2} \parens{ \braket{\an[m] [\hat{O},\ad[m]]} + \braket{[\an[m],\hat{O}]\ad[m]} }.
\label{eq:Mboson_EOM}
\end{equation}
Let us consider the case of $M = 2$, which is the minimal model for observing the power-law behavior. We denote the excitation number operator as $\hat{n} = \ad \an$. Setting $\hat{O} = \ad \an$ in Eq.~\eqref{eq:Mboson_EOM}, and using the commutation relation $[\an, \ad] = 1$, yields the equation of motion
\begin{equation}
    \frac{d\braket{\hat{n}}}{dt} = \kappa_1 + 4\kappa_2 + (-\gamma_1 + \kappa_1 + 2\gamma_2 + 6\kappa_2) \braket{\hat{n}} - 2(\gamma_2 - \kappa_2)\braket{\hat{n}^2}.
\end{equation}
At the critical point, i.e., $\gamma_2 = \kappa_2$, the mean excitation number $\braket{\hat{n}(t)}$ can be solved exactly to give
\begin{equation}
    \braket{\hat{n}(t)} = \braket{\hat{n}(0)} e^{-(\gamma_1 - \kappa_1 - 8\gamma_2)t} + \frac{\kappa_1 + 4\gamma_2}{\gamma_1 - \kappa_1 - 8\gamma_2} \sparens{1 - e^{-(\gamma_1 - \kappa_1 - 8\gamma_2)t}}.
\end{equation}
In terms of the power-law exponent
\begin{equation}
    \nu = \frac{\gamma_1 - \kappa_1}{4\gamma_2}
\end{equation}
from Eq.~\eqref{eq:powerlawexponent}, the mean excitation number can be written as
\begin{equation}
\boxed{
    \braket{\hat{n}(t)} = \braket{\hat{n}(0)} e^{-4(\nu - 2)\gamma_2 t} + \frac{1 + \kappa_1/4\gamma_2}{\nu - 2} \sparens{1 - e^{-4(\nu-2) \gamma_2 t}}.
}
\label{eq:mean_exact}
\end{equation}
Clearly, if $\nu < 2$, $\braket{\hat{n}(t)} \sim \exp[4(2-\nu)\gamma_2 t]$ diverges exponentially in time. Next, we set $\hat{O} = \hat{n}^2$. At the critical point, we have the equation of motion
\begin{equation}
    \frac{d\braket{\hat{n}^2}}{dt} = \kappa_1 + 8\gamma_2 + (\gamma_1 + 3 \kappa_1 + 16 \gamma_2) \braket{\hat{n}} + 2(-\gamma_1 + \kappa_1 + 12\gamma_2) \braket{\hat{n}^2}.
\end{equation}
The solution for $\braket{\hat{n}^2(t)}$ is given by
\begin{equation}
\boxed{
    \braket{\hat{n}^2(t)} = \braket{\hat{n}^2(0)} e^{-8(\nu - 3)\gamma_2 t} + \frac{1+\kappa_1/8\gamma_2}{\nu - 3} \sparens{1 - e^{-8(\nu - 3)}\gamma_2 t} + (\gamma_1 + 3 \kappa_1 + 16 \gamma_2) \int_0^{t} dt^\prime \braket{\hat{n}(t^\prime)} e^{8(\nu - 3)\gamma_2 (t^\prime - t)}.
}
\label{eq:secondmoment_exact}
\end{equation}
The integral in the final term can be evaluated explicitly by substituting the expression for $\braket{\hat{n}(t)}$ from Eq.~\eqref{eq:mean_exact}, although the resulting expression is rather complicated. By inspection, the expression has the form
\begin{equation}
    \braket{\hat{n}^2(t)} = \frac{3}{2} \frac{(1+\kappa_1/4\gamma_2)(1+\kappa_1/6\gamma_2)}{(\nu - 2)(\nu - 3)} + C_1 e^{-4(\nu - 2) \gamma_2 t} + C_2 e^{-8(\nu - 3)\gamma_2 t}
\end{equation}
for some coefficients $C_1$ and $C_2$. Thus, for $\nu < 3$, $\braket{\hat{n}^2(t)} \sim \exp[8(3-\nu)\gamma_2 t]$ diverges exponentially in time. The divergences in $\braket{\hat{n}(t)}$ and $\braket{\hat{n}^2(t)}$ are consistent with the power-law behavior.

We now derive a recurrence relation for the steady state expectation values $\braket{\ad[k] \an[k]}$, assuming $\nu > k+1$ such that the steady state values exist. These can be converted to steady state values for the moments $\braket{\hat{n}^k}$ of the number distribution.

Setting $\hat{O} = \ad[k] \an[k]$ in Eq.~\eqref{eq:Mboson_EOM} gives the equation of motion
\begin{equation}
\begin{split}
\frac{d\braket{\ad[k]\an[k]}}{dt} &= \frac{\gamma_1}{2} \parens{\braket{\ad[] [\ad[k]\an[k],a]} + \braket{[\ad,\ad[k]\an[k]]\an}} \\
&+ \frac{\kappa_1}{2} \parens{\braket{\an[][\ad[k]\an[k],\ad]} + \braket{[\an,\ad[k]\an[k]]\ad}} \\
&+ \frac{\gamma_2}{2}\parens{\braket{\ad[2] [\ad[k]\an[k],\an[2]]}+\braket{[\ad[2],\ad[k]\an[k]]\an[2]}} \\
&+ \frac{\kappa_2}{2} \parens{\braket{\an[2][\ad[k]\an[k],\ad[2]]} + \braket{[\an[2],\ad[k]\an[k]]\ad[2]}}.
\end{split}
\end{equation}
Using the identities $[\an,\ad[k]] = k \ad[k-1]$ and $[\an[2],\ad[k]] = 2k\ad[k-1]\an + k(k-1)\ad[k-2]$, which can be derived from the bosonic commutation relation, we obtain
\begin{equation}
\begin{split}
    \frac{d\braket{\ad[k]\an[k]}}{dt} &= k^2(k-1)^2 \kappa_2 \braket{\ad[k-2]\an[k-2]} + k^2(\kappa_1 + 4k \kappa_2)\braket{\ad[k-1]\an[k-1]} \\&- k\sparens{\gamma_1 - \kappa_1 - (k-1)(\gamma_2 - \kappa_2) - 4(k+1)\kappa_2} \braket{\ad[k]\an[k]} - 2k(\gamma_2 - \kappa_2) \braket{\ad[k+1]\an[k-1]}.
\end{split}
\end{equation}
In the steady state, this gives a recurrence relation for $\braket{\ad[k] \an[k]}$. At the critical point, $\kappa_2 = \gamma_2$, we have
\begin{equation}
\boxed{
    \braket{\ad[k]\an[k]} = \frac{k}{\nu - (k+1)}\sparens{\parens{k + \frac{\kappa_1}{4\gamma_2}}\braket{\ad[k-1]\an[k-1]} + \parens{\frac{k-1}{2}}^2 \braket{\ad[k-2]\an[k-2]}}
}
\label{eq:recurrence_moments}
\end{equation}
valid for $k \in \mathbb{Z}^+$ and $\nu > k+1$. As an application of this formula, we now derive asymptotic expressions for bosonic correlation functions at large $\nu$.

\subsection{Higher-order bosonic correlation function}
\label{app:correlation_func}
The $k$-th order bosonic correlation function (at zero delay) is defined as
\begin{equation}
    g^{(k)}(0) \equiv \frac{\braket{\ad[k]\an[k]}}{\braket{\ad\an}^k}
\label{eq:gk0_defn}
\end{equation}
evaluated at the steady state. If we regard the system as a photon source, $g^{(k)}(0)$ can be interpreted as a measure of the likelihood for $k$ photons to be emitted simultaneously, i.e., $k$-photon bunching.

Taking the limit $\nu \to \infty$, and using the recurrence relation~\eqref{eq:recurrence_moments}, we have the asymptotic expressions
\begin{equation}
\begin{split}
    &\braket{\ad \an} \sim \frac{1}{\nu}, \quad \braket{\ad[2]\an[2]} \sim \frac{1}{2\nu} \\ 
    &\braket{\ad[3]\an[3]} \sim \frac{15}{2\nu^2}, \quad \braket{\ad[4]\an[4]} \sim \frac{9}{2\nu^2} \\
    &\braket{\ad[5]\an[5]} \sim \frac{525}{2\nu^3}, \quad \braket{\ad[6]\an[6]} \sim \frac{675}{4\nu^3} \\
    &\braket{\ad[7]\an[7]} \sim \frac{99225}{4\nu^4}, \quad \braket{\ad[8]\an[8]} \sim \frac{33075}{2 \nu^4}.
\end{split}
\end{equation}
We can derive general asymptotic expressions for $\braket{\ad[k]\an[k]}$. Observe that for even $k$, the dominant contribution on the right hand side of Eq.~\eqref{eq:recurrence_moments} comes from $\braket{\ad[k-2]\an[k-2]}$. Therefore, setting $k = 2l$ we have the asymptotic recurrence relation for the even moments:
\begin{equation}
    \braket{\ad[2l] \an[2l]} \sim \frac{2l}{\nu} \parens{\frac{2l-1}{2}}^2 \braket{\ad[2l-2]\an[2l-2]},
\end{equation}
which can be solved to yield
\begin{equation}
    \boxed{\braket{\ad[2l]\an[2l]} \sim \frac{l!(2l-1)!!^2}{2^l \nu^l}.}
\label{eq:evenmoments_largenu}
\end{equation}
For odd $k$, the contributions from $\braket{\ad[k-1]\ad[k-1]}$ and $\braket{\ad[k-2]\an[k-2]}$ on the right hand side of Eq.~\eqref{eq:recurrence_moments} are comparable. Setting $k = 2l+1$, we have the asymptotic recurrence relation
\begin{equation}
    \braket{\ad[2l+1]\an[2l+1]} \sim \frac{2l+1}{\nu} \sparens{l^2 \braket{\ad[2l-1]\an[2l-1]} + \frac{l!(2l-1)!!^2 (2l+1)}{2^l \nu^l}}
\end{equation}
where we used Eq.~\eqref{eq:evenmoments_largenu}. This can be solved by making the ansatz $\braket{\ad[2l+1]\an[2l+1]}\sim f(l) \nu^{-(l+1)}$ for some function $f(l)$ to be determined, and the result is
\begin{equation}
    \boxed{\braket{\ad[2l+1]\an[2l+1]} \sim \frac{(l+1)!(2l+1)!!^2 (2l+3)}{3(l+1) 2^l \nu^{l+1}}.}
\end{equation}
Substituting the asymptotic expressions into the definition of $g^{(k)}(0)$~\eqref{eq:gk0_defn} gives 
\begin{equation}
    g^{(k)}(0) \propto \nu^{\lfloor k/2 \rfloor}
\end{equation}
for large $\nu$. This can potentially be used as an experimental signature of the power-law behavior.

\subsection{Wigner quasiprobability distribution}
\label{app:wigner}

As stated in the main text, we define the quadrature operators
\begin{equation}
    \hat{q} = \frac{1}{\sqrt{2}} \parens{\an + \ad}, \quad \hat{p} = -\frac{i}{\sqrt{2}} \parens{\an - \ad}
\label{eq:quadrature_ops}
\end{equation}
satisfying the canonical commutation relation $[\hat{q},\hat{p}] = i$. If the system physically describes a quantum particle trapped in a one-dimensional potential, $\hat{q}$ and $\hat{p}$ are the dimensionless position and momentum operators. 

The quantum state $\rho$ can be represented in phase space by the Wigner quasiprobability distribution $W(q,p)$~\cite{carmichael1999statistical,breuer2007theory,gardiner2000quantum}. For the $M$-boson model, the steady state Wigner distribution is rotationally symmetric. Thus, it is more convenient to work with $W(\alpha,\alpha^*)$, where $\alpha = (q+ip)/\sqrt{2}$ is the complex amplitude. The Lindblad master equation~\eqref{eq:Mboson_ME_supp} can be converted to an equation of motion for the Wigner distribution. To perform this conversion, it is helpful to recall the operator mapping dictionary~\cite{drummond2014quantum}
\begin{equation}
\begin{split}
    &\an \rho \longrightarrow \parens{\alpha + \frac{1}{2} \partial_{\alpha^*}} W(\alpha,\alpha^*) \equiv D_1 W(\alpha,\alpha^*), \quad \rho \ad \longrightarrow D_1^* W(\alpha,\alpha^*) \\
    &\rho \an \longrightarrow \parens{\alpha - \frac{1}{2}\partial_{\alpha^*}} W(\alpha,\alpha^*) \equiv D_2 W(\alpha,\alpha^*), \quad \ad \rho \longrightarrow D_2^* W(\alpha,\alpha^*)
\end{split}
\end{equation}
which maps the bosonic operators (which act in Hilbert space) to differential operators (which act in phase space), denoted by $D_1$ and $D_2$. It can be checked that the differential operators satisfy the commutation relations
\begin{equation}
    [D_1,D_1^*] = [D_2,D_2^*] = [D_1, D_2] = 0
\end{equation}
and
\begin{equation}
    [D_1,D_2^*] = 1.
\end{equation}
We can further exploit the rotational symmetry of the model by writing
\begin{equation}
    W(\alpha,\alpha^*) = W(r), \quad  r = |\alpha|,
\end{equation}
since the steady state Wigner distribution only depends on the radial coordinate $r = |\alpha|$. The Wigner distribution is normalized as follows:
\begin{equation}
    1 = \int d\alpha d\alpha^* \, W(\alpha,\alpha^*) = 2\pi \int_0^{\infty} dr \, r W(r).
\end{equation}
It is convenient to work with the coordinate $R \equiv r^2$, such that $W(R) \equiv W(r = \sqrt{R})$. We have the useful identities 
\begin{equation}
    \alpha \partial_\alpha = \alpha^* \partial_\alpha^* = R \partial_R, \quad \partial_\alpha \partial_\alpha^* = \partial_R R \partial_R.
\end{equation}
Using the operator mapping dictionary, one can show that the steady state Wigner distribution $W(R)$ satisfies the differential equation
\begin{equation}
\begin{aligned}
    0 = &(\gamma_1 - \kappa_1)(1+R\partial_R) W + \frac{\gamma_1+\kappa_1}{2} \partial_R R \partial_R W + (\gamma_2 - \kappa_2) \parens{4R + 2R^2 \partial_R + \frac{1}{2} R \partial_R^2 R \partial_R}W \\&+ 2(\gamma_2 + \kappa_2)(R \partial_R + R \partial_R R \partial_R)W.
\end{aligned}
\end{equation}
At the critical point $\kappa_2 = \gamma_2$, this simplifies to
\begin{equation}
    0 = (\gamma_1 - \kappa_1)(1+R\partial_R) W + \frac{\gamma_1+\kappa_1}{2} \partial_R R \partial_R W + 4\gamma_2 (R \partial_R + R \partial_R R \partial_R)W.
\end{equation}
Substituting the ansatz
\begin{equation}
    W(R) \propto \frac{1}{R^\nu},
\end{equation}
where $\nu$ is to be determined, we have
\begin{equation}
    0 = (\gamma_1 - \kappa_1) \frac{1 - \nu}{R^{\nu}} + \frac{\gamma_1 + \kappa_1}{2} \frac{\nu^2}{R^{\nu + 1}}+ \frac{4\gamma_2 \nu(\nu-1)}{R^{\nu}}.
\end{equation}
For large $R$, we obtain
\begin{equation}
    \nu \approx \frac{\gamma_1 - \kappa_1}{4 \gamma_2}
\end{equation}
which is exactly the power-law exponent~\eqref{eq:powerlawexponent} for the number distribution. The other solution $\nu = 1$ is unphysical, since it leads to non-normalizable Wigner distributions.

In terms of the phase space coordinates $q$ and $p$, we have the tail behavior
\begin{equation}
    W(q,p) \propto \frac{1}{(q^2 + p^2)^{\nu}},
\end{equation}
as stated in the main text. The probability distribution for the quadrature can be obtained by taking the marginal of $W(q,p)$. For example,
\begin{equation}
    P(q) = \int_{-\infty}^{\infty} dp \, W(q,p) \propto \frac{1}{q^{2\nu - 1}},
\end{equation}
and similarly for $P(p)$.

\section{Scaling behavior of the number statistics}
\label{app:scaling}
Here, we derive analytical predictions for the divergences of the various statistics presented in the main text, such as the mean number of excitation quanta and the second-order correlation function, as we approach the critical point $\delta \to 0^+$. While the numerical results shown in the main text are for the special case of $M=2$, the analytical treatment here holds for the general $M$-boson model.

To obtain the scaling behavior in terms of $\delta$, we can use the asymptotic behavior of $\rho_{n,n}$ derived in Eq.~\eqref{eq:truncatedpowerlaw}. However, the analysis can be made much simpler by considering the continuous probability distribution
\begin{equation}
    p(n) = \begin{cases} A n^{-\nu}, \quad &n \in \sparens{1,\frac{1}{\delta}} \\ 0, \quad &\text{else}
    \end{cases}
\end{equation}
where $A$ is a normalization factor. We regard $p(n)$ as a proxy for the discrete distribution $\rho_{n,n}$. The exponential cutoff for $\rho_{n,n}$ at $n \sim 1/\delta$ is replaced by a sharp cutoff for $p(n)$ at $n = 1/\delta$. The main idea is that the divergences in the number statistics can only arise from the power-law behavior of $\rho_{n,n}$. Thus, the scaling predictions from analyzing $p(n)$ will be qualitatively correct. Similarly, approximating the discrete distribution by a continuous distribution does not affect the qualitative nature of the divergences stemming from the power-law.

The normalization factor $A$ can be evaluated as
\begin{equation}
    A = 
    \begin{cases} 
    \dfrac{\nu - 1}{1 - \delta^{\nu - 1}}, \quad &\nu > 0 \, \,\text{and} \, \, \nu \neq 1 \\[2ex] \dfrac{1}{\log(1/\delta)}, \quad &\nu = 1
    \end{cases}
\end{equation}
which vanishes as $\delta^{1-\nu}$ for $0 < \nu < 1$ and is finite for $\nu > 1$. For $k \in \mathbb{Z}^+$, the $k$-th moment of the distribution $p(n)$ is given by
\begin{equation}
    \braket{n^k} = A \int_{1}^{1/\delta} dn \, n^{k-\nu} = \frac{A}{k-\nu + 1} \parens{\frac{1}{\delta^{k-\nu+1}} - 1}
\end{equation}
if $\nu \neq k+1$, and $\braket{n^k} = A \log(1/\delta)$ if $\nu = k+1$. This yields the scaling behavior (as $\delta \to 0^+$)
\begin{equation}
    \braket{n^k} \sim
    \begin{cases}
        \dfrac{1}{\delta^k}, \quad& 0 < \nu < 1 \\[2ex] 
        \dfrac{1}{\delta^k \log(1/\delta)}, \quad& \nu = 1 \\[2ex]
        \dfrac{1}{\delta^{k-\nu+1}}, \quad& 1 < \nu < k+1 \\[2ex]
        \log(1/\delta), \quad& \nu = k+1 \\[2ex]
        \text{finite}, \quad& \nu > k+1
    \end{cases}
\end{equation}
We can use this to derive the scaling behavior of the normalized second-order correlation function
\begin{equation}
    g^{(2)}(0) = \frac{\braket{n^2} - \braket{n}}{\braket{n}^2} \sim 
    \begin{cases}
        \text{finite}, \quad& 0 < \nu < 1 \\[2ex]
        \log(1/\delta), \quad& \nu = 1 \\[2ex]
        \dfrac{1}{\delta^{\nu - 1}}, \quad& 1 < \nu < 2 \\[2ex]
        \dfrac{1}{\delta \log^2(1/\delta)}, \quad& \nu = 2 \\[2ex]
        \dfrac{1}{\delta^{3-\nu}}, \quad& 2 < \nu < 3 \\[2ex]
        \log(1/\delta), \quad& \nu = 3 \\[2ex]
        \text{finite}, \quad& \nu > 3
    \end{cases}
\end{equation}
which is divergent in the regime $1 \leq \nu \leq 3$, as $\delta \to 0$. This is consistent with the numerical results shown in the main text.

\section{Quantum Li\'{e}nard system}
\label{app:quantum_lienard}

The quantum Li\'{e}nard system, proposed in Ref.~\cite{chia2025quantization}, serves as a quantum mechanical analog to the (classical) Li\'{e}nard system, which is well-studied in the context of nonlinear dynamics. The general Li\'{e}nard system is given by the dynamical equations
\begin{equation}
\begin{split}
    \dot{q} &= p, \\
    \dot{p} &= -f(q) p - g(q)
\end{split}
\label{eq:classical_lienard}
\end{equation}
defined on a two-dimensional phase space with coordinates $(q,p)$. Physically, $q$ and $p$ can be thought of as the position and momentum of a particle moving on a line. $f(q)$ and $g(q)$ are real and continuously differentiable functions. Li\'{e}nard's theorem~\cite{strogatz2024nonlinear} guarantees that the system has a unique and stable limit cycle surrounding the phase space origin if the following conditions are satisfied:
\begin{enumerate}
\item $g$ is an odd function, i.e., $g(-q) = -g(q)$.
\item $f$ is an even function, i.e., $f(-q) = f(q)$.
\item $g(q) > 0$ for $q > 0$.
\item $F(q) \equiv \int_0^q f(u) du$ has exactly one positive root at some $q = a$, negative for $0 < q < a$, positive and nondecreasing for $q > a$.
\end{enumerate}
Notable examples of dynamical systems that satisfy Li\'{e}nard's theorem include paradigmatic models such as the van der Pol, Rayleigh, and Duffing oscillators.

Inspired by Eq.~\eqref{eq:classical_lienard}, we consider the quantum equation of motion
\begin{equation}
\begin{split}
    \frac{d\braket{\hat{q}}}{dt} &= \braket{\hat{p}} \\
    \frac{d\braket{\hat{p}}}{dt} &= -\braket{:f(\hat{q})\hat{p}:} - \braket{:g(\hat{q}):}
\end{split}
\label{eq:quantum_lienard_general}
\end{equation}
where $\hat{q}$ and $\hat{p}$ are the quadrature operators from Eq.~\eqref{eq:quadrature_ops}. The notation $\braket{: \hat{O} :}$ indicates normal ordering, after expressing the operator $\hat{O}$ in terms of $\an$ and $\ad$. Here, we restrict to the case where $f$ and $g$ are at most cubic polynomials. In this restricted setting, the most general system that satisfies Li\'{e}nard's theorem, and thus guaranteed to be classically stable, has
\begin{equation}
    f(q) = -k_0 + k_2 q^2, \quad g(q) = k_1 q + k_3 q^3,
\label{eq:cubic_lienard}
\end{equation}
where $\{k_0,k_1,k_2,k_3\}$ are non-negative parameters. This corresponds to the model stated in the main text.

\subsection{Quantization of dissipative dynamical systems}
Eq.~\eqref{eq:quantum_lienard_general} cannot be simulated directly, since the system of equations is not closed in general. Instead, we seek a generator of infinitesimal time translation in the form of a Lindblad master equation. A general construction of the master equation (referred to as quantization) for an arbitrary polynomial $f$ and $g$ can be found in Ref.~\cite{chia2025quantization}. For completeness, we briefly summarize the steps involved in the quantization procedure:
\begin{enumerate}
    \item Start with the classical equation of motion of interest, of the form
    \begin{equation}
        \dot{q} = f_1(q,p), \quad \dot{p} = f_2(q,p),
    \end{equation}
    where $f_1$ and $f_2$ are arbitrary polynomial functions.
    \item Introduce the classical complex variable $\alpha = (q+ip)/\sqrt{2}$, which has the equation of motion
    \begin{equation}
        \dot{\alpha} = \frac{1}{\sqrt{2}} \sparens{f_1(\alpha,\alpha^*) + i f_2(\alpha,\alpha^*)},
    \end{equation}
    where $f_1(\alpha,\alpha^*) \equiv f_1(q=(\alpha+\alpha^*)/\sqrt{2}, p = -i(\alpha-\alpha^*)/\sqrt{2})$ (with a slight abuse of notation), and similarly defined for $f_2(\alpha,\alpha^*)$.
    \item Since $\alpha$ is a classical variable (sometimes called a $c$-number), we can change the order of $\alpha$ and $\alpha^*$ freely in $f_1(\alpha,\alpha^*)$ and $f_2(\alpha,\alpha^*)$. For convenience, reorder the terms such that all the $\alpha^*$ appear on the left of $\alpha$. 
    \item Making the substitution $\alpha \to \an$ and $\alpha^* \to \ad$, we can write down the quantum equation of motion
    \begin{equation}
        \frac{d\braket{\an}}{dt} = \frac{1}{\sqrt{2}} \sparens{\braket{:f_1(\an,\ad):} + i \braket{:f_2(\an,\ad):}}.
    \label{eq:general_quantumEOM}
    \end{equation}
The normal ordering of operators comes from Step 3. The choice of ordering here is simply for convenience. In principle, the master equation can be constructed for any operator ordering.
    \item Construct a Lindblad master equation
    \begin{equation}
        \dot{\rho} = -i[\hat{H},\rho] + \sum_{j} \Gamma_j \mathcal{D}[\hat{c}_j]\rho
    \end{equation}
    for a suitable (and in general non-unique) choice of Hamiltonian $\hat{H}$, jump operators $\hat{c}_j$ and decay rates $\Gamma_j > 0$, such that Eq.~\eqref{eq:general_quantumEOM} is satisfied. The solution to this inverse problem is the subject of Ref.~\cite{chia2025quantization}, which also proved the existence of a master equation for arbitrary polynomials $f_1$ and $f_2$. 
\end{enumerate}

\subsection{Example: Cubic Li\'{e}nard system}
The application of the quantization procedure to the cubic Li\'{e}nard system~\eqref{eq:cubic_lienard} is detailed in Ref.~\cite{chia2025quantization}. The result is the Lindblad master equation
\begin{equation}
\label{eq:quantum_lienard_ME}
    \dot{\rho} = -i[\hat{H},\rho] + k_0 \mathcal{D}[\ad]\rho + \frac{k_2}{2} \mathcal{D}[\ad \an - \ad[2]/2]\rho + \frac{3k_2}{8} \mathcal{D}[\an[2]]\rho,
\end{equation}
with the Hamiltonian
\begin{equation}
    \hat{H} = \hat{H}_0 + \hat{H}_1,
\end{equation}
\begin{equation}
    \hat{H}_0 = \frac{k_1+1}{2} \ad \an + \frac{3 k_3}{8} \ad[2] \an[2]
\end{equation}
and
\begin{equation}
    \hat{H}_1 = \frac{i}{4}\sparens{k_0 - i(k_1 - 1)} \an[2] - \frac{i}{4} \parens{\frac{k_2}{2} + i k_3} \ad \an[3] - \frac{i}{16}(k_2 + i k_3) \an[4] + \text{h.c.},
\end{equation}
where $\text{h.c.}$ denotes the Hermitian conjugate of all the preceding terms. As stated in the main text, the parameters $k_2$ and $k_3$ govern the nonlinear damping and anharmonicity of the oscillator, respectively. This can be seen from the classical equation of motion, or the above quantum master equation. Note that in the quantum case, the parameters $\{k_0,k_1,k_2,k_3\}$ all contribute to the (generalized) squeezing drive denoted by $\hat{H}_1$. This gives rise to quantum coherence in the steady state, i.e., off-diagonal elements of $\rho$ in the number basis.

In the main text, it was also stated that the power-law behavior cannot be observed for weak nonlinear dissipation. To see this, suppose $k_2$ is much smaller than all other parameters. Then, we can make the rotating-wave approximation on the master equation, which yields
\begin{equation}
\begin{split}
    \dot{\rho} &\approx -i[\hat{H}_0 + \hat{H}_1,\rho] + k_0 \mathcal{D}[\ad]\rho + \frac{k_2}{2} \mathcal{D}[\ad \an]\rho + \frac{3 k_2}{8} \mathcal{D}[\an[2]]\rho, \\
    \hat{H_1} &\approx \frac{i}{4}\sparens{k_0 - i(k_1-1)}\an[2] + \frac{k_3}{4} \ad \an[3] + \frac{k_3}{16} \an[4] + \text{h.c.},
\end{split}
\label{eq:RWA_quantumlienard}
\end{equation}
with $\hat{H}_0$ unchanged. Systematic corrections to the master equation in powers of $k_2$ can be calculated using the Krylov-Bogoliubov averaging~\cite{shen2023quantum}. 

The approximate master equation~\eqref{eq:RWA_quantumlienard} is reminiscent of the quantum Stuart-Landau model, studied in Refs.~\cite{walter2014quantum,sonar2018squeezing,mok2020synchronization,lorch2016genuine} in the context of quantum synchronization. Classically, the Stuart-Landau model describes the behavior of a system near a Hopf bifurcation~\cite{strogatz2024nonlinear}. In the quantum case~\eqref{eq:RWA_quantumlienard}, the steady state density matrix $\rho$ displays some signatures of a Hopf bifurcation (e.g., the emergence of a quantum limit cycle), but the number distribution vanishes exponentially.

\subsection{Additional numerical results}\label{app:lienard_additional_numerics}
\begin{figure}
    \centering
    \includegraphics[width=0.95\linewidth]{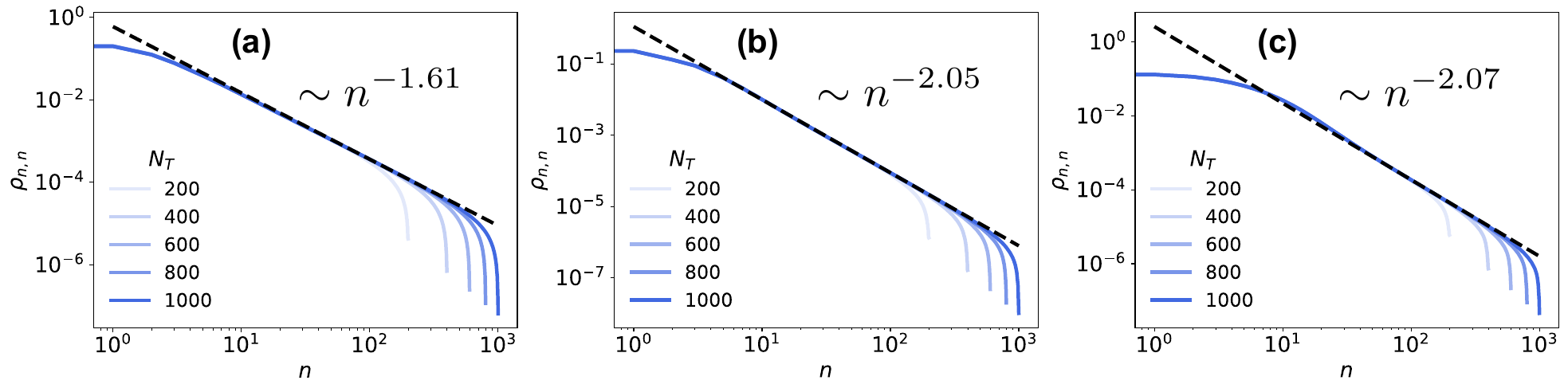}
    \caption{Steady state populations $\rho_{n,n}$ of the quantum Li\'enard system against the excitation number $n$. The Hilbert space truncation $N_T$ in the simulation ranges from $200$ to $1000$. The black dashed line indicates the power-law fit for $n \in [50,300]$ and $N_T = 1000$, with the fitted exponent indicated in the figures. The parameters $(k_0,k_1,k_2,k_3)$ used are: (a) (0.465, 0.353, 1, 0.042), (b) (0.168, 0.821, 1, 0.405), (c) (0.723, 0.372, 1, 0.788).}
    \label{fig:lienard_additional}
\end{figure}

As stated in the main text, the power-law behavior in the quantum Li\'enard model appears for typical parameters, where $k_2$ is not perturbatively small. Here, we present additional numerical evidence to support this claim. We numerically simulate the steady state of the Lindblad master equation~\eqref{eq:quantum_lienard_ME} using the Python package QuTiP~\cite{qutip5}. Since the bosonic Hilbert space is infinite-dimensional, we impose a truncation $N_T$ on the Fock basis $\{\ket{n}\}$ in the numerical simulation, up to $N_T = 1000$. The steady state distribution $\rho_{n,n}$ is plotted in Fig.~\ref{fig:lienard_additional} for various parameters. From the numerical data, we find that the power-law behavior clearly emerges for some range of $n$, and appears to fall off rapidly near $n = N_T$ (similar behavior holds for the coherences, not shown here). We believe this is a numerical artifact of the truncation, and not a genuine cutoff as we had for the $M$-boson model away from the critical point. We validate this by varying $N_T$ and observing that the location of the `cutoff' grows proportional to $N_T$. Therefore, we posit that the asymptotic behavior of $\rho_{n,n}$ obeys the power law. It is, however, possible that the quantum Li\'enard system does have a true cutoff at some large $n$ way beyond what we simulated. Nonetheless, we expect the power-law behavior to be valid over a wide range of $n$, which can be physically observed.

\begin{figure}
    \centering
    \includegraphics[width=0.8\linewidth]{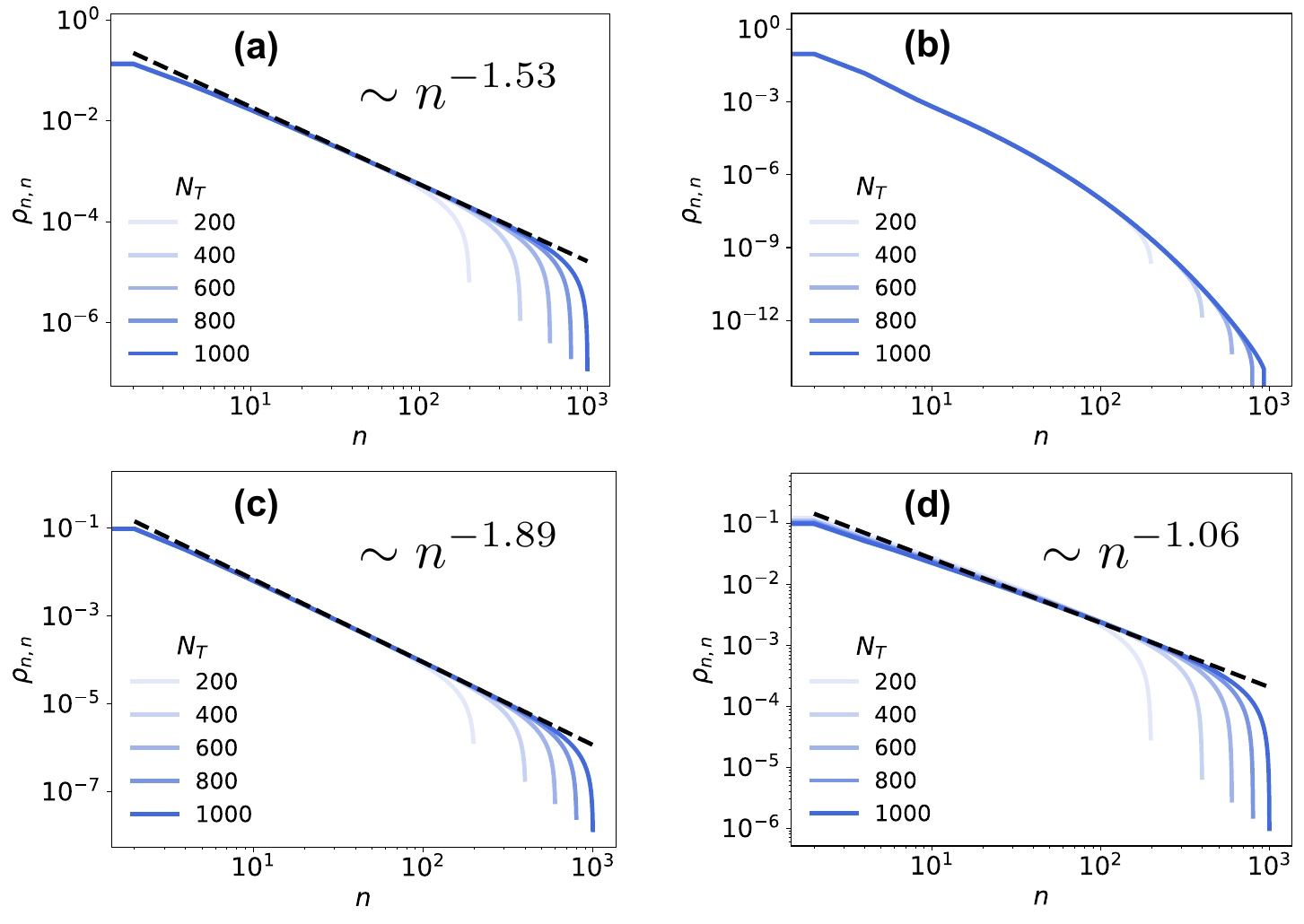}
    \caption{Steady state populations $\rho_{n,n}$ against the excitation number $n$, in the even parity sector. For odd values of $n$, $\rho_{n,n} = 0$ which are omitted from the plots. The Hilbert space truncation $N_T$ in the simulation ranges from $200$ to $1000$. The black dashed line indicates the power-law fit for $n \in [50,300]$ and $N_T = 1000$, with the fitted exponent indicated in the figures. The systems studied are: (a) Quantum Li\'{e}nard model with $k_0 = k_1 = k_3 = 0$ and $k_2 = 1$. (b) Same as (a), but with the Hamiltonian $H = 0$. (c) Same as (a), but with the Hamiltonian $H = -i(\ad \an[3] - \ad[3] \an)/4$. (d) Same as (a), but with the Hamiltonian $H = -i(\an[4] - \ad[4])/8$.}
    \label{fig:lienard_robustness}
\end{figure}

\dmtwo{Next, we study the robustness of the power-law behavior in the quantum Li\'{e}nard model. The numerical results are shown in Fig.~\ref{fig:lienard_robustness}. As explained in the main text, we identify the nonlinear dissipation term governed by $k_2$ to be the key ingredient for the power law. For example, the power law tail can be observed even when $k_0 = k_1 = k_3 = 0$ and $k_2 = 1$ [Fig.~\ref{fig:lienard_robustness}(a)]. Now, if we artificially set the Hamiltonian $H$ to zero, we get a simplified master equation
\begin{equation}
    \dot{\rho} = -i[H,\rho] + \frac{1}{2} \mathcal{D}[\ad \an - \ad[2]/2]\rho + \frac{3}{8} \mathcal{D}[\an[2]]\rho
\end{equation}
with $H = 0$. In this case, the steady state does not exhibit a power law tail [Fig.~\ref{fig:lienard_robustness}(b)]. We observe that the power law can be recovered by setting $H = -i(\ad \an[3] - \ad[3] \an)/4$ [Fig.~\ref{fig:lienard_robustness}(c)] or $H = -i(\an[4] - \ad[4])/8$ [Fig.~\ref{fig:lienard_robustness}(d)]. In other words, this suggests that not all the nonlinear terms in the master equation for the quantum Li\'{e}nard model are needed for the power-law behavior.

Note that in all the models studied in Fig.~\ref{fig:lienard_robustness}, the steady state has a two-fold degeneracy corresponding to the $\mathbb{Z}_2$ symmetry $\hat{a} \to -\hat{a}$. The results are shown for the steady state in the even parity sector, with similar results observed for the steady state in the odd parity sector.
}

\end{document}